\documentclass[]{report}

\usepackage[dvips]{graphics}

\renewcommand{\abstract}[1]{{ \footnotesize \noindent {\bf
Abstract} #1 \\}}
\renewcommand{\author}[1]{\subsection*{#1}}
\newcommand{\address}[1]{\subsection*{\it#1}}

\begin{document}

\chapter*{ULTRA HIGH ENERGY COSMIC RAYS: present status and future prospects}

\author{A A Watson}


\address{Department of Physics and Astronomy,
University of Leeds,
Leeds LS2 9JT, UK}

\abstract{  Reasons for the current interest in cosmic rays above 
$10^{19}$ eV are
described.  The latest results on the energy spectrum, arrival direction distribution and mass composition of cosmic rays are reviewed,
including data that were reported after
the meeting in Blois in June 2001.  The enigma set by the existence
of ultra high-energy cosmic rays remains.  Ideas proposed to explain it
are discussed and progress 
with the construction of the Pierre Auger Observatory is outlined.} 

\section{Introduction}For the purposes of this review I define
ultra high-energy cosmic rays (UHECRs) as those cosmic rays having 
energies above $10^{19}$ eV. There is 
currently great interest in them, partly because we have little
idea as to how Nature creates particles or photons of these 
energies.  Also we know enough about their 
energy spectrum and arrival direction distribution to believe that
we have an additional problem: their sources must be reasonably 
nearby (within 100 Mpc) but 
there is no evidence of the anisotropies anticipated
if the galactic and inter-galactic 
magnetic fields are as weak as astronomers tell us.

The distance limit comes from a combination of well-understood
particle physics and the universality of the 2.7 K radiation.  
Interactions of protons and heavier nuclei with 
this, and other, radiation fields degrade the energy of particles rather rapidly.  In the case of protons, the reaction is photopion 
production, while heavier nuclei are 
photodisintegrated by the 2.7 K radiation and the diffuse infrared background.  These effects were first recognised by Greisen, and
by Zatsepin and Kuzmin, and lead to the expectation that the energy spectrum 
of cosmic rays should terminate rather sharply above $4 \times 10^{19}$ eV 
(the GZK cut-off).  Above $4 \times 10^{19}$ eV about 50\% of 
particles must come from within 130 Mpc, while at 
$10^{20}$ eV the corresponding distance is 20 Mpc.

The most recent data suggest that particles do exist with energies beyond the GZK cut-off and that the arrival direction distribution
is isotropic.  The mass of the cosmic rays above $10^{19}$ eV is 
not known, although there are experimental limits on the fraction of 
photons that constrain one of the models proposed to resolve the enigma.

\section{Measurement of UHECR}  The properties of UHECRs are obtained by studying 
the cascades, or extensive air showers (EAS), they create in the atmosphere.  
Many methods of observing these cascades have been explored.  
Currently two approaches seem to be most effective.  In one, the density 
pattern of particles striking an array of detectors laid out on the ground 
is used to infer the primary energy.  At $10^{19}$ eV the
 footprint of the EAS on the ground is several square kilometres so 
detectors can be spaced many hundreds of metres apart.  Alternatively, on 
clear moonless nights, the fluorescence light emitted when shower particles excite 
nitrogen molecules in the atmosphere can be observed by massive photomultiplier 
cameras.  This technique, uniquely, allows the rise and fall of the cascade in the atmosphere to be inferred.

The primary energy of the initiating particle or photon is deduced in 
different ways. For the detector arrays, Monte Carlo calculations have shown 
that the particle density at distances from 400 -- 1200 m is closely proportional 
to the primary energy.  Such a density can be measured accurately (usually 
to around 20\%) and the primary energy inferred from parameters found by 
calculation.  The estimate of the energy depends on the realism of the 
representation of features of particle interactions within the 
Monte Carlo model, at energies well above accelerator energies.  The currently 
favoured model (QGSJET) is based on QCD and is matched to accelerator 
measurements.  Although this model appears to describe a variety of data 
from TeV energies up to $10^{20}$ eV \cite{nagano1}, one cannot be certain 
of the systematic error in the energy estimates.

For the fluorescence detectors, the primary energy is found by 
integrating the number of electrons in the cascade curve and assuming that 
their rate of energy loss is close to that at the minimum of the dE/dx curve for 
electrons, $\sim2.2$ MeV per g cm$^{-2}$ in the case of air.  
A small model-dependent correction must be 
made to account for the energy carried by muons and neutrinos into the 
ground.  Ideally, one wants to 
compare estimates of the primary energy made in the same shower by the two 
techniques operating simultaneously, but this has yet to be done at these 
energies.  So far all that has been possible is to compare estimates 
of the fluxes at nominally the same energy. 

\section{ The Energy Spectrum, Arrival Direction Distribution and Mass of UHECRs}  
\subsection{Energy Spectrum}    At the time of the Blois meeting (June 2001) 
data on the energy spectrum from a number of experiments had
seemed in good accord \cite{nagano2}.  In particular, the rates of events 
at $10^{19}$ eV reported by different experiments were in agreement at 
the 10 -- 15\% level.  In addition, preliminary data 
from the Utah-based HiRes group, reported at the International 
Cosmic Ray Conference in 1999, showed 7 events above $10^{20}$ eV, 
in good agreement with the number anticipated from the flux reported 
by the Japanese AGASA array.  

The situation has now changed dramatically.  At the international 
meeting in Hamburg (August 2001), the AGASA group reported 
additional data, quite consistent with their earlier work, and described 17 
events above $10^{20}$ eV.  The HiRes group 
reported on monocular data obtained with one of their cameras 
from an exposure slightly greater than that of the AGASA group.  
Assuming a spectrum similar to that reported earlier by 
the AGASA group, the HiRes team 
had expected to see about 20 events above $10^{20}$ eV, but observed only 2.  
This unexpected discrepancy is not yet understood.

\begin{figure*}
\begin{center}
\scalebox{0.45}{\includegraphics{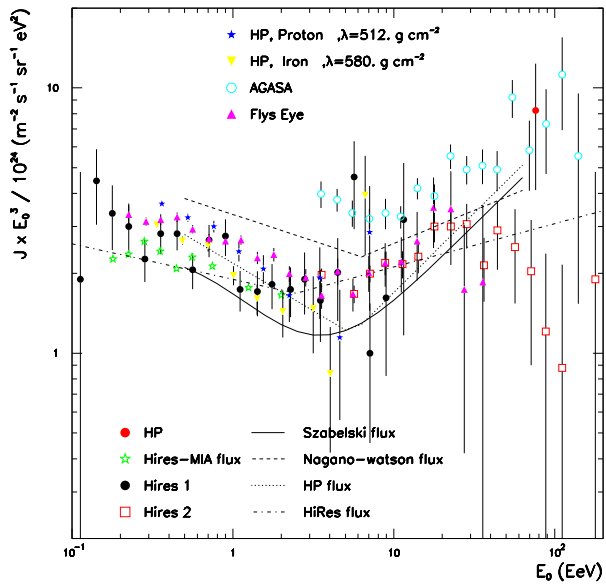}}
\end{center}
\caption{A composite energy spectrum from AGASA, Fly's Eye, Haverah Park and 
HiRes.  This plot was prepared with the help of Maximo Ave.}
\label{fig:cs}
\end{figure*}

The data from Fly's Eye (the earliest fluorescence 
experiment), Haverah Park (a ground array that used water-Cherenkov 
detectors), HiRes and AGASA are shown in figure~\ref{fig:cs}.  There are several
points to note.  The Haverah Park energy estimates have been 
re-assessed \cite{ave1} using the QGSJET model.  
In the range $3 \times 10^{17}$ to $3 \times 10^{18}$ eV there is very 
good agreement between the Fly's Eye, Haverah Park and HiRes results.  
A recent Haverah Park analysis \cite{ave2} 
suggests that protons and iron are in the ratio 35:65 in 
this energy range.  With this mixture the spectrum agreement is even better.
%
%
This implies that the QGSJET 
model provides an adequate description of important features of 
showers up to $10^{18}$ eV.  However, the AGASA energies have also 
been estimated with the QGSJET model under the assumption that the 
primaries are protons at energies above $3 \times 10^{18}$ 
eV, the lowest AGASA energy plotted.  There is no evidence as 
to what mass species is dominant at the highest energies but the methods used 
would lead to an estimate lower by only about 20\% if iron nuclei were 
assumed.  This change would be 
insufficient to reconcile the AGASA-HiRes differences, particularly 
with regard to the point at which the spectrum slope flattens 
above $10^{18}$ eV.  However a combination of a change in the QGSJET 
model and iron primaries (for which there is no evidence) 
might go some way to aligning the different results at the highest energies.

There are also unanswered questions about the HiRes data.  The 
`disappearance' of the events reported as being above $10^{20}$ eV in 1999 is 
attributed to a better understanding of the atmosphere which is now 
claimed to be clearer than had previously been supposed.  
The Hamburg results were prepared using an `average 
atmosphere' so presumably subsequently some events will be assigned larger 
energies and some smaller ones.  Two further issues need resolving.  
Firstly, an accelerator-based calibration of the fluorescence 
yield \cite{kakimoto} led to the claim `that the fluorescence 
yield of air between 300 and 400 nm is proportional to the electron 
dE/dx.'  This claim is not consistent with information tabulated in 
the paper, where it is shown that the yield from 50 keV electrons is 
similar to that from 1.4 MeV electrons.  Also
the dE/dx curve plotted there, normalised to the 1.4 MeV measurements, 
does not fit the accelerator data for 300, 
650 and 1000 MeV electrons.  The latter discrepancy is about 
15 -- 20\% and in such a direction as would increase the HiRes 
energies.  Secondly, Nagano et al. \cite{nagano3} has described a new 
measurement of the yield in air from 1.4 MeV electrons.  
In what seems to be a very careful study, they find that the earlier 
results \cite{kakimoto} gave a higher yield at 356.3 nm and 391.9 nm than is 
found now.  Nagano attributes the absence of background corrections 
as being responsible for at least some of the discrepancies 
\cite{nagano4}.  The longer wavelengths become increasingly important when 
showers are observed at the large distances common at the highest 
energies because of Rayleigh scattering.  The magnitude of the adjustments 
that need to be made to the HiRes data are presently 
unclear and further fluorescence yield measurements are 
certainly required.

At the Hamburg meeting, the HiRes group also reported data from their 
stereo system.  With 20\% of the monocular exposure, they found 
1 event with an energy estimated as being close to $3 \times 10^{20}$ eV, the 
energy of the largest event found with the Fly's Eye 
detector \cite{bird}.  My opinion is that the spectra from AGASA and 
HiRes will come together as further understanding is gained 
of the models and of the atmosphere.  
Knowledge of the mass composition will also help considerably.  For now 
it seems that trans-GZK events do exist but that the flux of 
them is less certain than appeared a few years ago.

\subsection{Arrival Direction Distribution} 
\begin{figure*}
\begin{center}
\scalebox{0.4}{\includegraphics{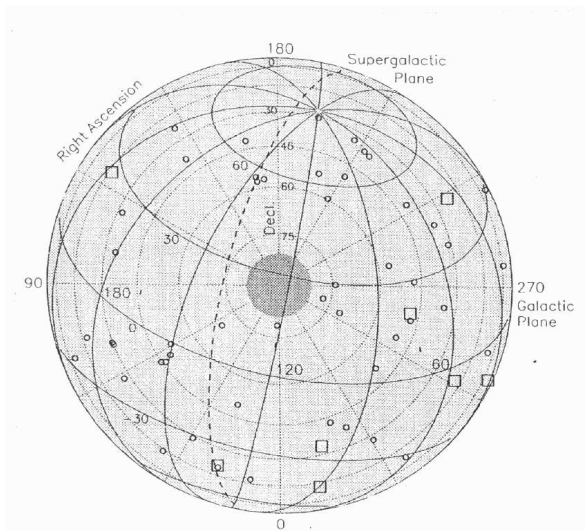}}
\end{center}
\caption{AGASA arrival direction distribution for 59 events above 
$4 \times 10^{19}$ eV.  The most energetic events ($>10^{20}$ eV) 
are shown by squares [20].}
\label{fig:ad}
\end{figure*}

The angular resolution of shower 
arrays and of fluorescence detectors is typically 2 -- 3$^{\circ}$.  
The arrival direction of the 59 events with 
energy above $4 \times 10^{19}$ eV registered by the AGASA 
group is shown in figure~\ref{fig:ad} 
\cite{takeda}.  The distribution is isotropic and 
there is no preference for events to come from close to the 
galactic or the super-galactic planes.  
The AGASA group draw attention to a number of clusters, where a cluster 
is defined as a grouping of 2 or more events within $2.5^{\circ}$.  
It is claimed that the number of doublets (5) and triplets (1) could have arisen 
by chance, with probabilities of 0.1\% and 1\%.  The implications of such 
clusters would be profound but the 
case for them is not proven.  The angular bin was not defined 
a priori and the data set used to make the initial claim for clusters is also 
being used in the `hypothesis 
testing' phase.  Furthermore, I note that the directions of the 7 
most energetic events observed by Fly's Eye, Haverah Park, 
Yakutsk and Volcano Ranch do not line up 
with any of the 6 cluster directions.  

%
It is hard to understand the isotropy if the local extragalactic 
magnetic field is really just $10^{-9}$ gauss.  A proton of $10^{20}$ eV 
would be deflected by only about $2^{\circ}$ over a distance 
of 20 Mpc if the field has a 1 Mpc correlation length \cite{kronberg}.  
If the fields were much higher, as has been suggested \cite{farrar}, 
then the lack of anisotropy might be understood, but more energy 
is then stored in the magnetic field and this 
might create other difficulties.  Similarly, if the charge of the 
particles initiating the showers was much higher than Z=1, 
the isotropy could be explained.

\subsection{Mass Composition} Interpretation of the data on UHECRs is 
hampered by our lack of knowledge of the mass of the incoming particles.  
Data from several experiments can be interpreted 
as indicating a change from a dominantly iron beam near $3 \times 10^{17}$ eV 
to a dominantly proton beam at $10^{19}$ eV.  But the situation is 
unclear and quite open at higher energies. The data are just too limited 
and the interpretations are ambiguous as 
both the fluorescence detectors and ground arrays rely 
on shower models to deduce composition information.  

It is 
unlikely that the majority of the events claimed to be near 
$10^{20}$ eV have photons as parents as some of 
the showers have the normal numbers of muons (the tracers of 
primaries that are nuclei) and the profile of the 
most energetic fluorescence event is inconsistent 
with that of a photon primary \cite{halzen}.  Furthermore, there is now 
evidence that less than 40\% of the events at $10^{19}$ eV are 
photon-initiated.  This limit has been set in two ways.  
Taking the energy spectrum as measured by Fly's Eye as being independent 
of the mass of the incoming particles, the rate of showers 
coming at large angles to the vertical can 
be calculated.  Using Haverah Park data, it has been found that the 
observed rate of inclined showers is much higher than would 
be expected if the primary particles were mainly photons \cite{ave3}.  
A more traditional 
attack on the problem by the AGASA group, searching for showers 
which have significantly fewer muons than normal, has 
given the same  upper limit \cite{shinosaki}.    
It is unlikely that many events are created 
by neutrinos as the distribution of zenith 
angles would be different from that observed.  Indeed, in all aspects so 
far measured, events of $10^{20}$ eV look like events of 
$10^{19}$ eV, but ten times larger, and this can be 
reiterated as we go to lower and lower energies were nuclei 
seem certain to be the progenitors of showers.

\section{Theoretical Interpretations} The UHECR enigma is attracting 
significant theoretical attention. Some ideas suppose 
a form of electromagnetic acceleration while others invoke new physics.  

Currently it is popularly believed 
that cosmic rays with energies up to about $10^{15}$ eV are energised 
by a process known as `diffusive shock acceleration'. Supernovae 
explosions are identified as the likely sites, although 
so far there is no direct evidence for acceleration of nuclei by 
supernova remnants at any energy.  The diffusive shock process, which has 
its roots in some early ideas of Fermi, has been extensively studied since 
its conception in the late 1970s.  In \cite{drury} it is shown that the 
maximum energy attainable is given by E = kZeBR$\beta$c, 
where B is the magnetic field in the region of the shock, R is 
the size of the shock region and k is a constant less than 1.  The same 
result has been obtained by a number of people, e.g. \cite{hillas}, and 
most authors agree upon it.  However, some claim that 
the diffusive shock acceleration process can be modified to give much 
higher energies than indicated by the equation and that radio galaxy 
lobes, in particular, are probable acceleration sites.  It is difficult 
to see how an energy of $3 \times 10^{20}$ eV can be accounted 
for if the size of the shock region is 10 kpc and the magnetic field 
is 10 $\mu$G (values thought typical of lobes of radio galaxies), 
as even the optimum estimate of the energy is lower by a factor 
of 3 than the observational upper limit.  It could be that the 
magnetic fields are stronger than is usually supposed, a line 
of argument that also comes from the arrival direction work mentioned above.

Proposals have 
been made which dispense with the need for electromagnetic acceleration.  
Attention has usually been focused on the highest energy 
events ($> 10^{20}$ eV).  However, it is my 
view that proposers of some of the more exotic mechanisms often 
overlook one or more important points.  Any mechanism able to explain the 
highest energy events must also explain those above about 
$3 \times 10^{18}$ eV, where the galactic component 
possibly disappears.  The spectrum above this point is possibly too 
smooth to imagine that there are two or more radically different 
components --- although this might be seen as an almost philosophical 
argument, particularly in the light of figure~\ref{fig:cs}!  
In addition, the solutions proposed must produce particles at the 
top of the atmosphere that can generate showers of the type we 
see and now understand rather well.  Finally, source energetics cannot be 
ignored: there seems little point in inventing a mechanism to `solve' the 
GZK cut-off problem that requires a source 
region that is unrealistically energetic. 

An overview of the various non-electromagnetic processes proposed can be found 
in \cite{nagano2} and I will only discuss one of these here.  
It has been suggested that UHECR arise from the decay of super-heavy relic 
particles.  In this picture, the cold dark matter is 
supposed to contain a small admixture of long-lived super-heavy 
particles with a mass $>10^{12}$ GeV and a lifetime 
greater than the age of the Universe \cite{berezinsky}.  
It is argued that such particles can be produced 
during reheating following inflation or through the decay of hybrid 
topological defects such as monopoles connected by 
strings.  I find it hard to judge how realistic these ideas are but 
the decay cascade from a particular candidate \cite{benalki} 
has been studied in some detail \cite{birkel} and 
\cite{rubin}.  A feature of the decay cascade is that an accompanying 
flux of photons and neutrinos is predicted which may be 
detectable with a large enough installation.
In particular photons are expected to be between 2 and 10 times
as numerous as protons above $10^{19}$ eV.   
The anisotropy question has been examined and 
specific predictions have been made for the anisotropy that would 
be seen by a Southern Hemisphere observatory.  
Observation of the predicted anisotropy, plus the 
identification of appropriate numbers of neutrinos and photons, would 
be suggestive of a super-heavy relic origin.  However,
the experimental 
results on the photon/proton ratio at $10^{19}$ eV described above
clearly do not support it.

\section{Detectors of the Future}
The Pierre Auger 
Observatory was conceived to measure the properties of the highest 
energy cosmic rays with unprecedented statistical precision.  When completed, 
it will consist of two instruments, in 
the Northern and Southern Hemispheres, each covering 
3000 km$^{2}$.  The design calls for a hybrid system with 
1600 particle detectors and three 
or four fluorescence detectors at each of the sites.  The particle 
detectors will be deep water-Cherenkov tanks arranged on a 1.5 km 
hexagonal grid.   These detectors have been selected because water acts 
as a very effective absorber of the multitude of low 
energy electrons and photons found at distances of about 1 km from the 
shower axis.  

At the Southern site fluorescence detectors 
will be set up at four locations, 
one near the centre of the particle array with the others on small 
promontories at the array edge: the site is close to the town of 
Malarg\"{u}e in Mendoza Province, Argentina.  During clear moonless nights, 
signals will be recorded in both the fluorescence detectors and the 
particle detectors, while for roughly 90\% of the time only 
particle detector data will be available.  
Construction of an engineering array 
containing 40 water tanks and a section of a fluorescence 
detector has been completed (September 2001) and all of the sub-systems of 
the Observatory have been demonstrated.  The first 
`hybrid' events were recorded in December 2001 and there is great 
confidence that the Observatory will work as designed.  When the 
Auger Observatory at Malarg\"{u}e has operated for 10 years, we would expect 
to have recorded over 300 events above $10^{20}$ eV.

Achieving an exposure greater than that 
targeted by the Auger Observatory is a formid\-able challenge.  
A promising line is the development of an idea due to Linsley.  The concept 
is to observe fluorescence light produced by showers from 
space with satellite-borne equipment.  It is proposed to 
monitor $\sim10^{5}$ km$^{2}$ sr (after allowing for an estimated 8\% on-time).  
Preliminary design studies have been carried 
out in Italy and the USA.  An Italian-led collaboration has proposed a 
design that is under study for flight in the International Space Station.  
This is known as EUSO (the Extreme Universe Space Observatory), and has 
the potential to detect neutrinos in large 
numbers as well as UHECRs.  Observations are scheduled to start in 
2008.  This type of project requires considerable 
technological development but may be the only way to push to energies 
beyond whatever energy limits are found with the Auger instruments.\\
\\  
{\bf Acknowledgements:}  I am grateful to the organisers for 
inviting me to the Blois meeting.  On-going support of PPARC to work 
on ultra high-energy cosmic rays at the University 
of Leeds is gratefully acknowledged.  I also thank 
my many colleagues in the Pierre Auger project 
for helping to make a 10-year-old dream 
become a reality.

\end{document}